\documentclass[aps,amsmath,twocolumn]{revtex4}
\usepackage{graphicx}

\begin{document}


\title{A Farewell to Falsifiability}

\author{Ali Frolop}\thanks{With help from Douglas Scott}
\altaffiliation{{\tt afrolop@phas.ubc.ca};
 Dept.\ of Physics \& Astronomy, University of British Columbia, Vancouver,
 Canada}
\author{Ali}
\altaffiliation{Ali Narimani; {\tt anariman@phas.ubc.ca};
 Dept.\ of Physics \& Astronomy, University of British Columbia,
 Vancouver, Canada}
\author{Frolov}
\altaffiliation{Andrei Frolov; {\tt frolov@sfu.ca}, Dept.\ of Physics,
 Simon Fraser University, Burnaby, BC, Canada}

\date{1st April 2015}

\begin{abstract}
Some of the most obviously correct physical theories -- namely string theory
and the multiverse -- make no testable predictions, leading many to
question whether we should accept something as scientific even if it makes no
testable predictions and hence is not
refutable.  However, some far-thinking physicists have proposed instead that
we should give up on the notion of Falsifiability itself.  We endorse this
suggestion but think it does not go nearly far enough.  We believe that we
should also dispense with other outdated ideas, such as Fidelity, Frugality,
Factuality and other ``F'' words.  And we quote a lot of famous people to
support this view.
\end{abstract}


\maketitle

\date{today}

\noindent
\underline{\sl Foundations}\quad
The evolution of scientific ideas is governed by several guiding principles.
However, these principles can themselves change with time, as the tangled
landscape of scientific knowledge develops.  The question of what makes
``good science''
and what principles we should be using to determine this, is probably most
obviously exemplified in the physical sciences through debates about the
nature of string theory and the multiverse.

Because of the obvious appeal of these theoretical ideas, despite the lack of
observable predictions, it has been suggested that it is time to retire one
of the long-cherished gold standards of science, namely {\it falsifiability\/}
\cite{Carroll}.

Although some scientists may see this as extreme, we believe it does not go
nearly far enough.  It is in fact time to retire several other scientific
principles as we strive to accept the truth.

\smallskip
\noindent
\underline{\sl Fundamentals}\quad
String theory and its close cousin, the notion of a multiverse, can solve all
of the existing problems in theoretical physics \cite{Wilczek}.
These include combining
gravity with quantum mechanics, explaining the values of all the physical
constants (including why the cosmological constant is at least $10^{60}$
times too small), and solving many other fundamental mysteries \cite{tangled}.
It has become popular to attack these ideas for making no testable
predictions \cite{doubters}.
However, although not widely discussed among physicists,
there have in fact been published studies demonstrating experimental
verification of the ideas of string theory \cite{string_theory}.  Moreover, the
nature of physical reality itself, and the existence of all the known particles
and their interactions, is surely proof enough.

The multiverse, too, has been criticised for its lack of
useful predictions.  However,
the idea that one should explain some of the mysteries of modern physics by
inventing an infinite number of entire universes seems quite natural
to us.  Although some accuse the proponents of the multiverse of being mere
purveyors of science fiction \cite{Moorcock}, that is simply
small-mindedness.  Instead, we would suggest that the great advantages of
extending the Universe into the multiverse should encourage us to think even
bigger than the multiverse \cite{perverse}.  Indeed, even a cursory search
shows that this concept is being widely discussed in the
literature \cite{comics}, as people come to terms with what to call a
collection of multiverses \cite{omniverse}.

\smallskip
\noindent
\underline{\sl Falsifiabilit}{\sl y}\quad
Karl Popper \cite{karl}
is usually credited with introducing the idea of falsifiability
as an important demarcation criterion for deciding what is scientific and
what is merely something else.  However, Popper's work first appeared in
the 1930s.  Are we to
suppose that any ideas before this period were incapable of being proved to
be scientifically sound, because falsifiability was yet to be invented?
The answer is obvious ``no'',
because Popper already knew which theories were correct, and so he
invented this idea of testability to shore up some of the shakier theories
of his time.  However, now that the merits of the string-multiverse are so
self-evident, we feel confident that if Popper were alive today, he would
never have introduced the obsolete notion of Falsifiability in the first place.

One of the examples that Popper gave for an endeavour that is demonstrably
unscientific is astrology, which he thought to be entirely unpredictive
\cite{astrology}.  However, as Thomas Kuhn pointed
out \cite{thomas}, the case is quite the opposite. 
Astrology {\it is\/} falsifiable, and there is nothing magic about 
this demarcation criterion.  However, Kuhn thought that science and scientists
are mostly concerned with finding and solving puzzles about the world, which 
is of course not true as we explain shortly.

Moreover, the idea of falsifiability has always had a basic flaw.  This
is an obvious problem of the
``strange loop'' sort \cite{GEB}, i.e., if we take Proposition A to be
``All correct theories contain falsifiability'', but then add Proposition B,
which states ``Proposition A is itself a theory'', then
we realise that we're in trouble.

We therefore reject the requirement of Falsifiability.  But there is more.

\smallskip
\noindent
\underline{\sl Other Fs}\quad
Several other notions are usually considered to be inviolable by practicing
scientists.  However, we
consider these also to be oversimplifications whose time has now come to be
challenged -- they are spherical sacred cows, if you like.

One such idea is reliability or repeatability, which we might call
{\it fidelity}.
We have come to expect that a good scientific theory should give the same
answer each time a question is asked.  For example, ``Will the Sun come up
tomorrow?'' should have the same comforting response each time.  However,
string theory contains a much richer phenomenology.  The landscape
of string vacua contains something like $10^{500}$ possibilities.  Therefore
the experiment of making a universe will give a different answer each time.
And so we see that we should also give up Fidelity.

Another idea is simplicity \cite{simplicity}, or theoretical {\it frugality},
sometimes referred to as Ockham's Razor.
Carroll \cite{Lewis}, the assassin of falsifiability, also suggests that the
more preposterous the universe the better \cite{preposterous}.
General Relativity (GR) is patently not simple, and string theory is to GR as
a tax form is to a simple yes/no question.
But given the
enormous successes of string theory, it becomes clear that the best scientific
theories should be as complicated as possible.  Mathematical complexity and
unfathomability should be criteria to apply when deciding what makes
good science.  In the same way, the idea of the multiverse
is the very opposite of economy, proposing a fecundity of universes
instead of just one parsimonious sphere of existence.  On top of that,
string theory requires many more physical dimensions than are accessible to
the human senses, and yet there can be no doubt that these exist.
Hence we should also give up on Frugality.

The idea that lies at the core of most thinking 
about science, including both Kuhn and Popper, is that the ideas 
should not only be testable, but should pass those tests and hence be 
actually correct.  This idea, which we might call {\it factuality},
has influenced scientists so much that Feynman 
once said ``I think it's much more interesting to
live not knowing than to have answers which might be wrong'' \cite{richard}.
However, this has been misinterpreted, since the Universe -- as we know it --
is just a tiny piece of a vast entity described by the stringy-verse.
What the father of modern physics meant is that we should not be obsessed
with things being correct.  As a famous philosopher of the 20th century
said ``I'd far rather be happy than right any day'' \cite{douglas}.
Hence we should seek out theories that are
not even correct within the small piece of the multiverse where we happen to
live.  This enables us to free ourselves from the need to explain empirical
data at all.  Einstein once said ``Reality is merely an illusion, albeit a
very persistent one'' \cite{albert}.
A statement often attributed to Fermi is that the poorest sort of theory
is ``not even wrong''.  So let us embrace that, reject the notion that
a theory has to be correct and say goodbye to Factuality.

Freed from the fetters of rightness, string theory and the multiverse instead
become matters of pure belief.  This then is their ultimate triumph.  By
becoming faith-based science they reconcile those two great human
endeavours -- the ``non-overlapping magisteria'' of Stephen Jay Gould
-- bringing together science and religion.  As Einstein said ``I assert that
the cosmic religious experience is the strongest and noblest driving force
behind scientific research'' \cite{albert2}.


\smallskip
\noindent
\underline{\sl Finish}\quad
Ed Witten once called string theory ``a bit of 21st century physics that
somehow dropped into the 20th century''.  Presumably he was suggesting that
string theory came to us from the future, but we think that a much more
likely solution is that it came here from another part of the multiverse.
Perhaps, then, there are other universes out there in which string theory
is not only simple and correct, but even falsifiable as well.



\smallskip



\begin{thebibliography}{9}
\bibitem{perverse} We could take the Latin root `per-', meaning `beyond' or
 `completely', and combine it with the suffix `-verse'.
\bibitem{comics} One example arises in answers to questions like ``what would
happen if Spiderman fought Superman?'', since this involves combining the
Marvel Multiverse with the DC Multiverse.
\bibitem{astrology} However, Popper's skepticism towards astrology is just what
you'd expect from a Leo.
\bibitem{Lewis} And we mean Sean Carroll here, not the mathematician Lewis, who
once said ``I've believed as many as six impossible things before breakfast''.
\bibitem{albert} Not really relevant, but we realised that we hadn't quoted
Einstein yet.
\bibitem{albert2} Actually, that's a better quote, so ignore the last one.
\bibitem{further} Comments on this paper should be sent to the lead
 author, Dr.~Frolop.
\end{thebibliography}
\end{document}